          [2001/04/25 1.1 (PWD)]
\documentclass[a4paper,twocolumn]{esapub2005} 
\usepackage{times}
\usepackage{natbib}
\usepackage{graphicx}

\title{ Pulse profiles and cyclotron line energy dependence on X-ray pulsars luminosity}
\author{S. Tsygankov}
\author{A. Lutovinov}
\author{E. Churazov}
\author{R. Sunyaev}
\affil{Space Research Institute, Profsoyuznaya str. 84/32, Moscow 117997, Russia}

\begin{document}


\maketitle

\begin{abstract}
We present the results of broad band (3-100 keV) observations of several
X-ray pulsars with the INTEGRAL and RXTE observatories. We concentrate
on the luminosity and energy dependence of the pulse profile and
the variations of the cyclotron line energy. In V0332+53 the line
energy changes nearly linearly with the source luminosity, while in 4U0115+63
its behavior is more complicated. Strong variations of the pulse
profile with the energy and source intensity were found for
both of pulsars; in V0332+53 the
changes of the pulse profile near the cyclotron line are especially
drastic. The preliminary results obtained for Her X-1 and GX 301-2 in a high intensity
state show the absence of significant pulse profile changes with the energy. 
Results and possible emission mechanisms are briefly discussed in
terms of theoretical models of accreting pulsars.
\end{abstract}

\section{Introduction}

In this work two transient X-ray pulsars, V0332+53 and 4U0115+63, and
two persistent ones, Her X-1 and GX 301-2, were studied.
First two sources are members of high
mass X-ray binary systems with $Be$ class stars and demonstrate regular powerful outbursts in
which their luminosities can reach $several\times10^{38}$ erg s$^{-1}$.
The distinctive peculiarity of both sources is that not
only the main harmonics of the cyclotron resonance scattering feature (CRSF) are
registered in their X-ray spectra but also their higher harmonics. Another two
X-ray pulsars belong to different classes: GX 301-2 -- is a system with the 
supergiant companion, Her X-1 -- is a low-mass X-ray binary sistem. Spectra of both   
sources also include CRSFs (see, e.g., \cite{cob02}).

\section{Cyclotron line energy}

\subsection{V0332+53}

The pulsar V0332+53 was observed during a powerful outburst in 2004-2005 with the
range of measured luminosities from
$\sim10^{37}$ to $\sim5\times10^{38}$ erg s$^{-1}$.
The significant number of V0332+53 observations carried out with the
INTEGRAL and RXTE observatories allowed us to
reconstruct the source spectrum at different phases of the outburst
and to trace the evolution of its parameters in detail (\cite{tsyg06a}).

As a whole, the pulsar spectrum during the outburst
can be well described by a power law with an exponential
cutoff at high energies modified by several harmonics of CRSF
(see, e.g., \cite{kreyk05}; \cite{pots05}) that is observed for
several X-ray pulsars.
But the behaviour of the cyclotron line in this source deserves
a special attention, as its position is not a constant.
The line energy dependence on the source
luminosity obtained from \textit{INTEGRAL} and \textit{RXTE} data is
shown in Fig.\,1 by dark triangles and squares,
respectively. The formal
fitting of this dependence with a linear relation gives
$E_{cycl,1}\simeq-0.10L_{37}+28.97$ keV, where $L_{37}$ -- the source luminosity
in units of $10^{37}$ erg s$^{-1}$. Assuming that for low
luminosities the emission come practically from the neutron star
surface (see below) we can estimate the magnetic field on the
surface
$$B_{NS}=\frac{1}{\sqrt{1-\frac{2GM_{NS}}{R_{NS}c^2}}}
\frac{28.97}{11.6}\simeq3.0\times10^{12} {\rm G}   \qquad \qquad (1)$$
where $R_{NS}$ and $M_{NS}$ -- are the neutron star radius and mass,
respectively.

\begin{figure}
\centering
\includegraphics[width=0.95\columnwidth, bb=15 270 520 690,clip]{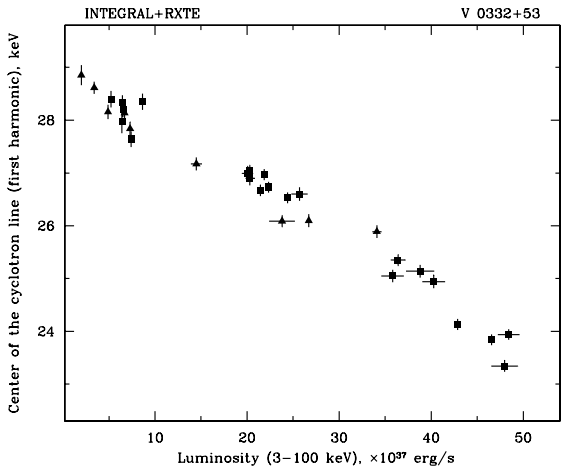}
\caption{The cyclotron line energy dependence on the source luminosity (3-100 keV) for V0332+52.
Triangles are \textit{INTEGRAL} results, squares are \textit{RXTE}
ones. }
\end{figure}

It was shown in \cite{basko76a} that there is a critical value of the
luminosity ($L^*\sim10^{37}$ erg s$^{-1}$) dividing two accretion
regimes: the regime when the influence of the radiation on the falling
matter is negligible and the regime when this influence is
significant. When $L<L^*$, the matter free-fall zone is extended
almost down to the surface of the neutron star. In the opposite case
($L>L^*$), observed for V0332+53, the radiation-dominated shock rises
high above the neutron star surface. Almost all of the kinetic energy
of the infalling gas is lost in this shock, and is then emitted
laterally by the sides of the accretion column.

\cite{basko76a} and \cite{lyubar88} showed that the height of the shock $H$
changes practically linearly with changing of the accretion rate,  $H\propto\dot m$,
in a wide range of values $\dot m$, i.e. the shock height grows linearly when
the source luminosity increases and can reach
several kilometres for high luminosities.

In case of V0332+53 the maximum relative change of the line energy and,
consequently, the corresponding magnetic field is about
$\sim25$\%. In an approximation of a dipole field of the neutron star,
it corresponds to a $7.5$\% relative change of the height $h$ where
the feature is formed. At the end of the outburst, the source
luminosity falls to $\sim10^{37}$ erg s$^{-1}$, 
the column height decreases and we detect the emission coming virtually
from the neutron star surface. Taking the neutron star radius $R_{NS} \sim 10^6$ cm
we can estimate the maximum height $h\sim 750$ m, that is much less than the shock height $H$
expected for such luminosities.

Due to the fact that only a small fraction of the energy
accumulated in the accreting matter is emitted at the shock
(\cite{basko76a}) and its
main part goes into the extended sinking zone below the shock
the registered emission is a
superposition of emissions from different heights above the neutron
star surface. Therefore, we can consider the height $h$ as a some
averaged or ``effective'' height of the formation of the cyclotron
feature not coinciding with the position of the shock itself.

\begin{figure}
\centerline{\includegraphics[width=0.95\columnwidth, bb=30 280 560 690,clip]{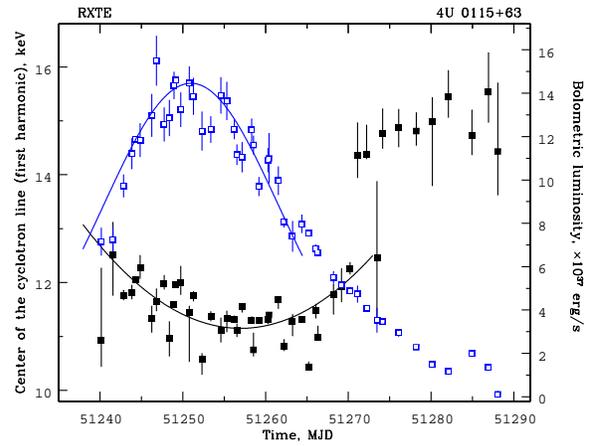}}
\caption{The cyclotron line energy (solid squares) and source luminosity (open squares) time dependence
obtained during the outburst of 4U0115+63. Solid lines represents best fit Gaussians.}
\end{figure}

\subsection{4U0115+63}

The spectrum of 4U0115+63 also can be well fitted by the power law model with an exponential
cutoff at high energies and it also contain CRSF with higher harmonics.
The energy of this feature  depends on the source luminosity,
but the picture of these changes is not as clear as in case of V0332+53.
The cyclotron line energy and source luminosity time dependences
obtained during the outburst of 4U0115+63 in Feb-Apr 1999 (see \cite{nak06} 
and references there) are presented in Fig.\,2.
Both observed dependences can be roughly divided in two parts:
where these dependences can be approximated by smooth law (high luminosity state, between MJD 51240 and 51270)
and where they become nearly linear or even constant for the cyclotron energy 
(below $\sim6\times10^{37}$ erg s$^{-1}$, after MJD 51270).
The physical reasons of such a behaviour will discussed in details in future (\cite{tsyg06b}).
Assuming that for the low luminosity state
the cyclotron feature is formed near the neutron star surface than we can estimate the magnetic field of
the neutron star using the equation (1), as $B_{NS}\simeq1.5\times10^{12} {\rm G}$.

\section{Pulse profile}

The observed strong spectral and geometrical changes of accretion column might lead to significant
changes in other observable characteristics of pulsars, for example, in the shape of the
pulse profile.

\subsection{V0332+53}

\begin{figure*}
\centerline{\hbox{
\includegraphics[width=0.69\columnwidth, bb=15 50 305 307, clip]{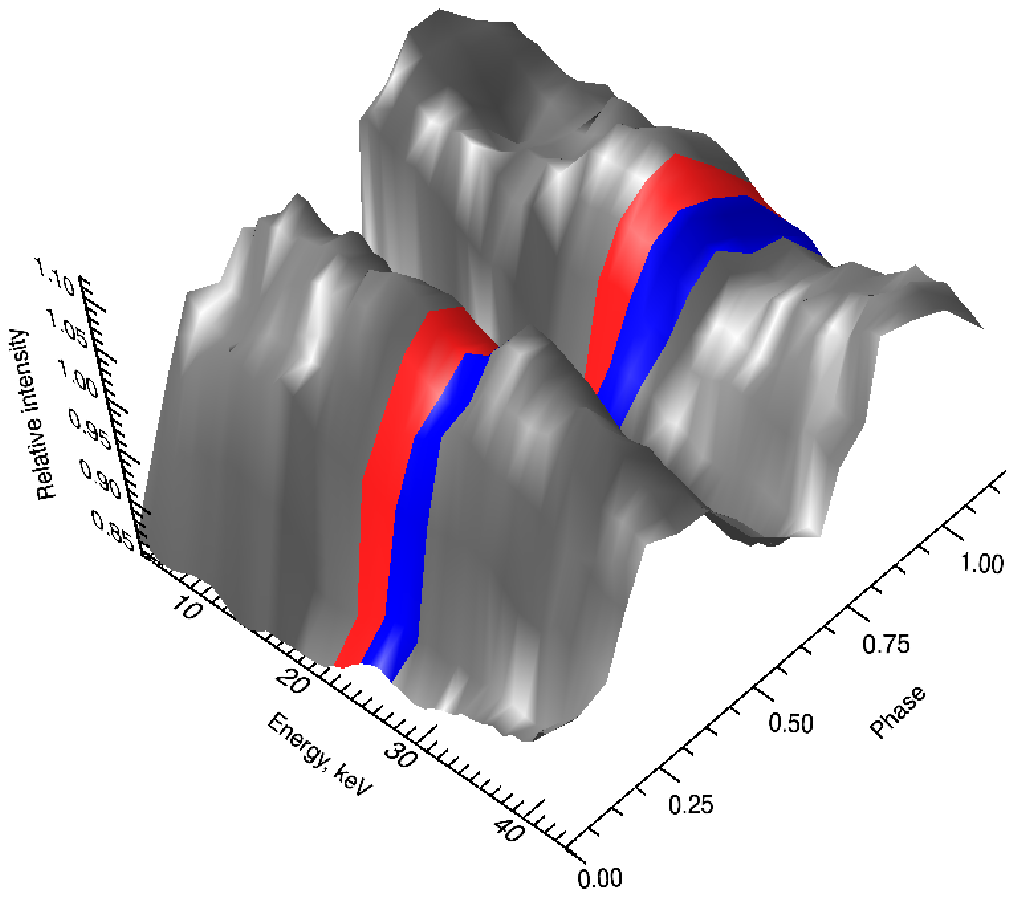}
\includegraphics[width=0.7\columnwidth, bb=25 70 291 280, clip]{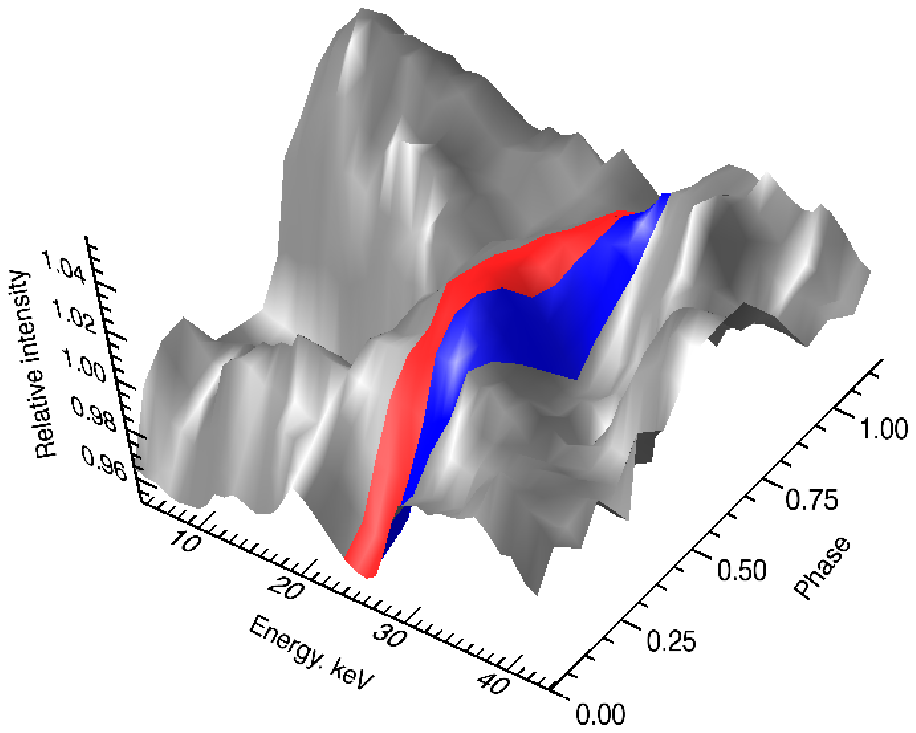}
}}
\centerline{\hbox{
\includegraphics[width=0.66\columnwidth, bb=40 300 530 740, clip]{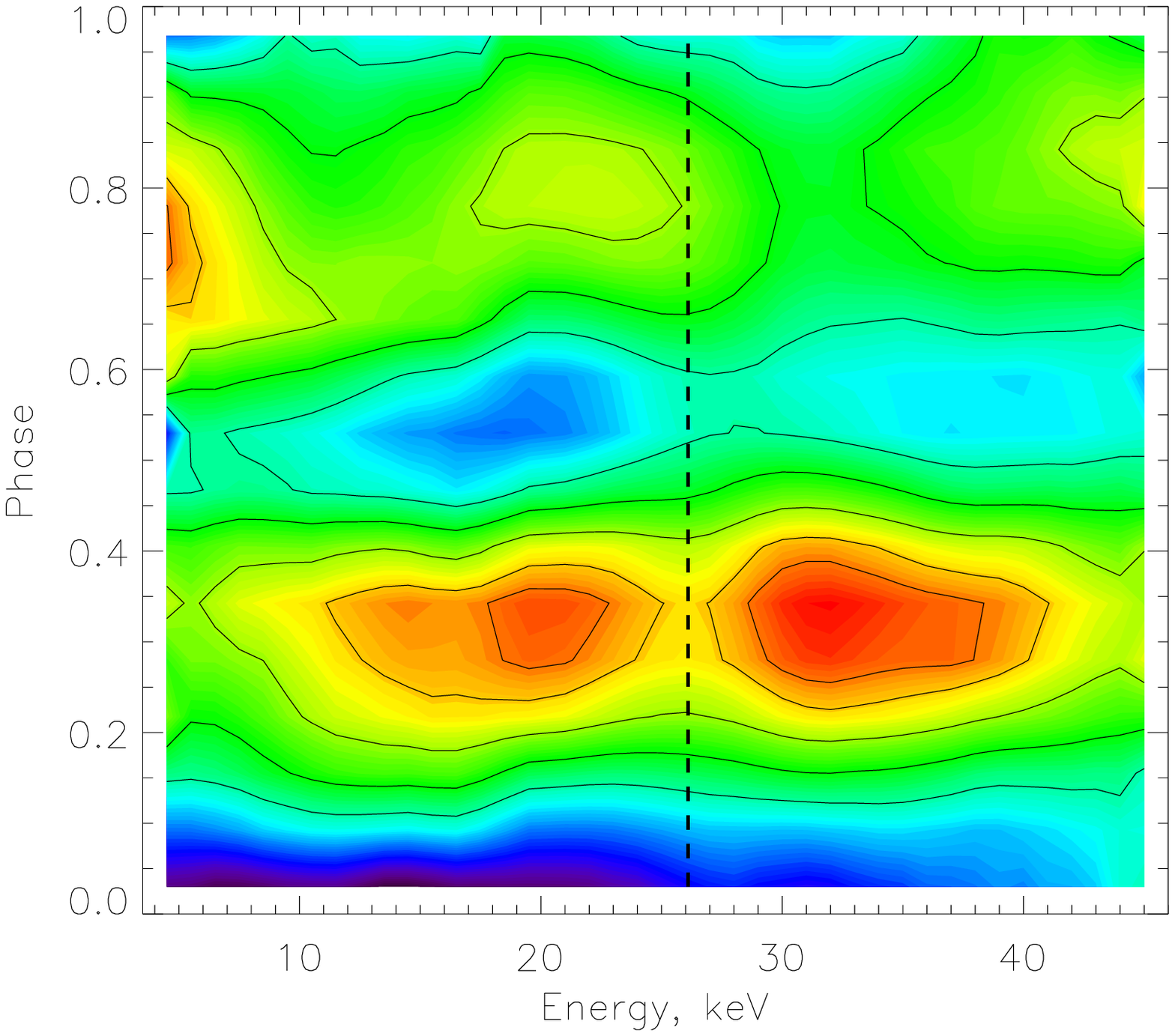}
\hspace{3mm}\includegraphics[width=0.69\columnwidth, bb=20 300 530 740, clip]{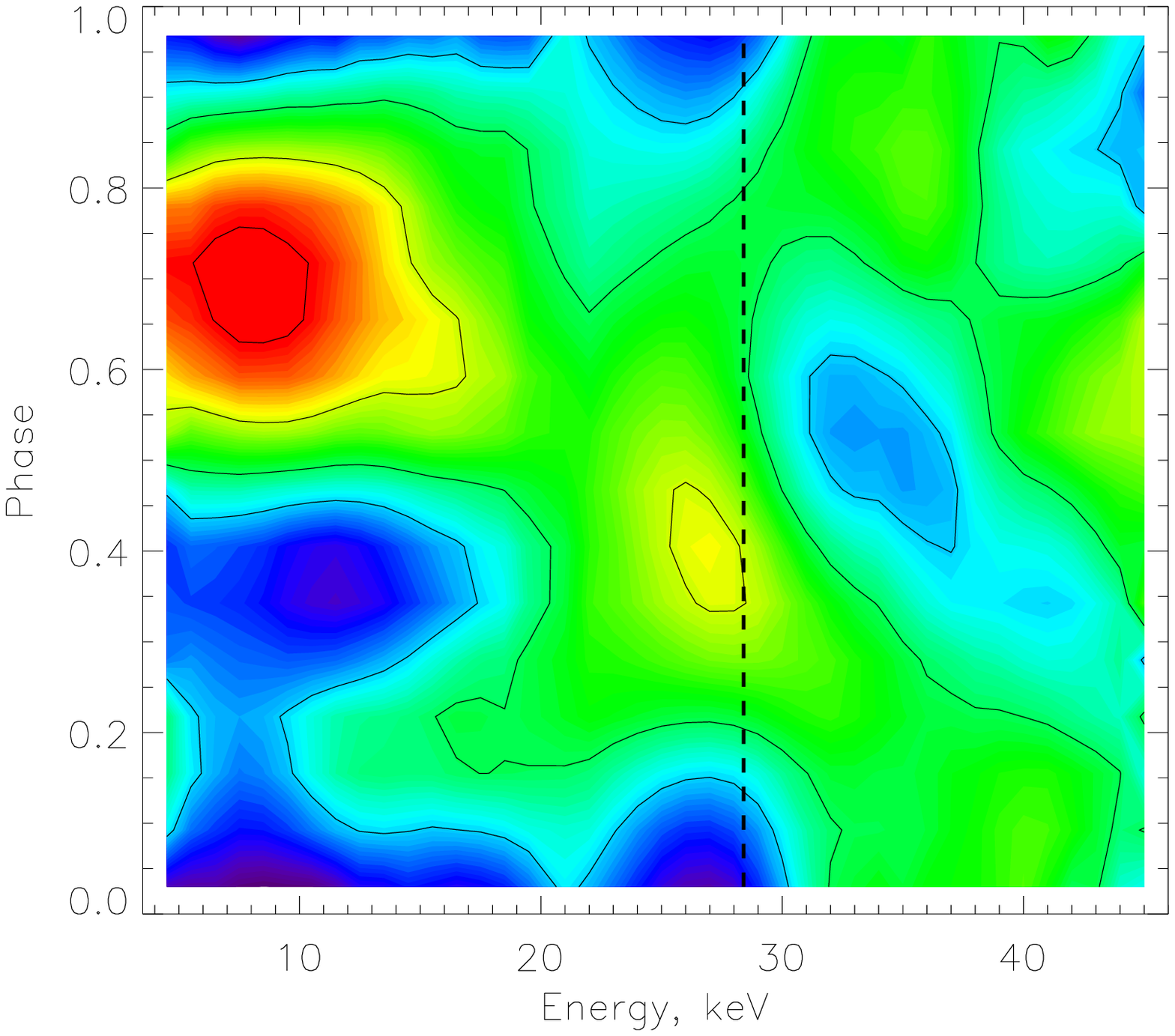}
}}
\caption{3D evolution of the pulse profiles (top) and 2D-distributions of profile intensities
(bottom) for different luminosities.  Positions of the cyclotron line center are shown by dashed
lines. Left part of the figure corresponds to the mean luminosities $\sim3.4\times10^{38}$ erg s$^{-1}$
and right one -- $\sim7.3\times10^{37}$ erg s$^{-1}$. }
\end{figure*}

Two three-dimensional pulse profiles (distributions of  pulse intensities along the pulse phase and
energy relative to the mean value)  for different states of V0332+53, $\sim3.4\times10^{38}$ (left part)
and $\sim7.3\times10^{37}$ erg s$^{-1}$ (right part), are shown in Fig.\,3 (upper
panel).
The red and blue stripes represents regions of lower and upper
wings of cyclotron lines. In bottom panels of Fig.\,3
two-dimensional distributions of pulse profile intensities are
demonstrated by different colors and levels of equal intensities. It
is interesting to trace changes of the maximum intensities for both
observations: in the high state the positions of both peaks are
practically unchanged with the energy, although their relative intensity 
is changed; in the low state the profile
became single-peaked at energies just below the cyclotron line
with a drastic transition to the double-peaked just above the line
energy (Fig.\,3).

The exact explanation of the pulse profile is complicated task, many geometrical and physical
processes can affect to the pulse profile changes. For example, the gas accretion stream flowing along
the Alfven surface covers only a part of this surface. Spinning with the same angular velocity as the
neutron star it can periodically shade from the observer different parts of emission regions, that can
explain the different intensity of peaks in the pulse profile and their changes with the energy
(\cite{basko76b}).

\begin{figure*}
\centerline{\includegraphics[angle=90,width=14cm,bb=55 60 565 775]{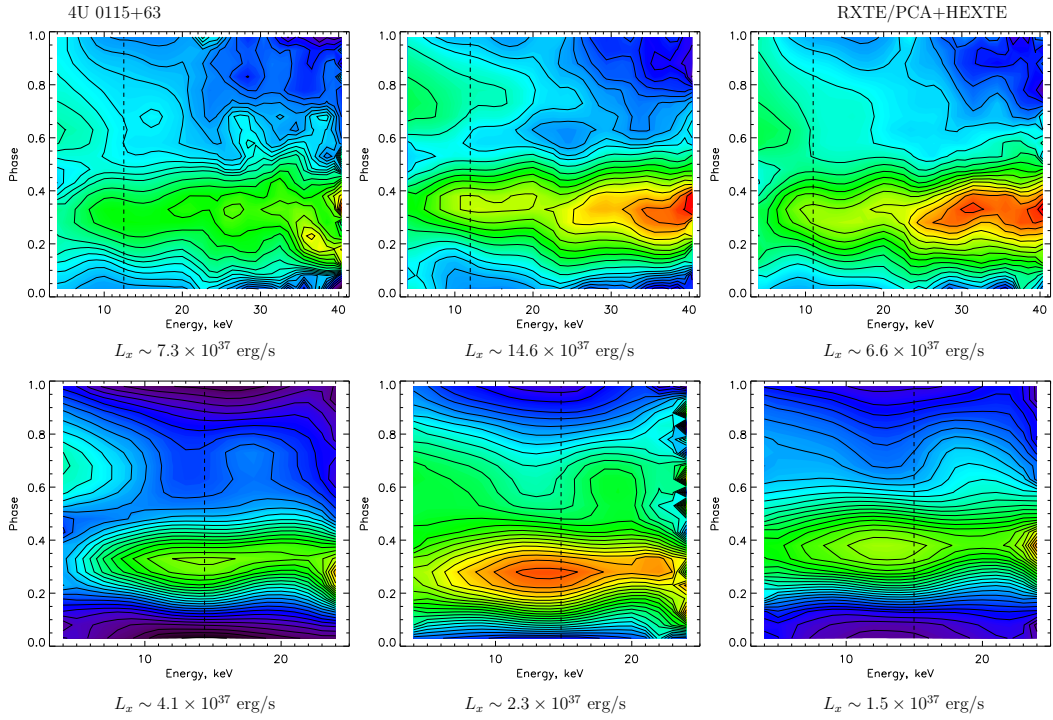}}
\caption{2D-distributions of profile intensities
for different source 4U0115+63 luminosities (3-100 keV).  Positions of the cyclotron line center are shown by dashed
lines. At the three last maps (for low luminosity cases) only PCA data is shown.}
\end{figure*}

\begin{figure*}
\centerline{\hbox{
\includegraphics[width=0.9\columnwidth, bb=40 300 530 740, clip]{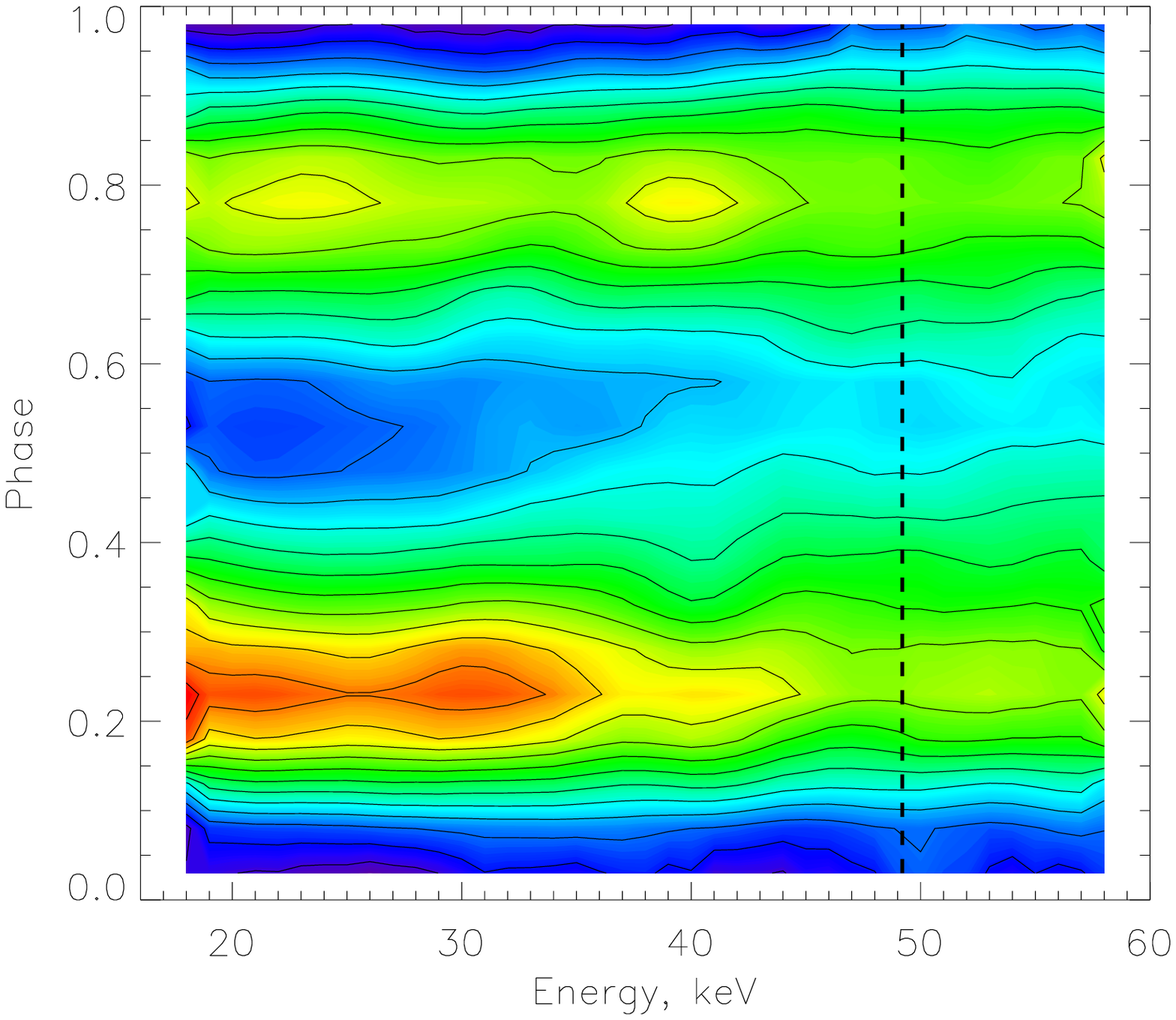}
\hspace{3mm}\includegraphics[width=0.9\columnwidth, bb=40 300 530 740, clip]{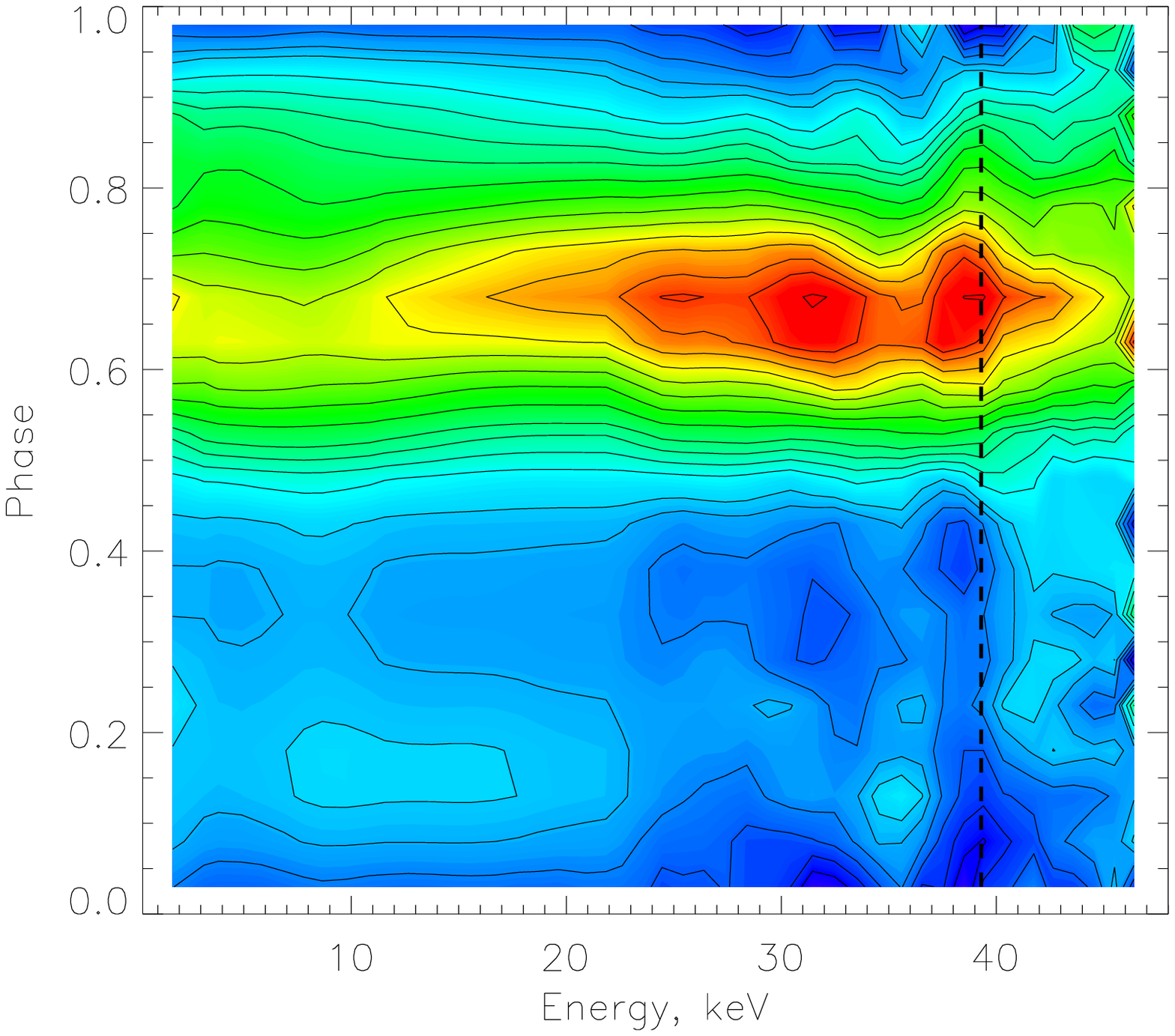}
}}
\caption{2D-distributions of profile intensities for GX 301-2 (left panel) and Her X-1 (right panel) at high 
intensity state.  Positions of the cyclotron line center are shown by dashed lines.}
\end{figure*}

The observed changes of the source pulse profile near the cyclotron frequency is
difficult to describe in detail within the framework of current
models. Naturally, they can be connected with peculiarities of the
radiation beaming near the cyclotron frequency (\cite{gnedin73}; \cite{pavl85}). In particular, as e.g.
\cite{meszar85} showed, the cyclotron line shape demonstrates a
strong angular dependence. The plasma is more transparent at large
angles than at small ones for energies below and above the line
energy. Therefore photons will escape predominantly in the directions
of large angles, i.e. the radiation beaming in different energy
channels near the cyclotron line will be strongly different.
But the exact physical picture is unclear yet.

\subsection{4U0115+63}

Two-dimensional distributions of pulse profile intensities
for 4U0115+63 obtained with the RXTE observatory are shown in Fig.\,4. In the bright state the profile is
double-peaked up to $\sim10-20$ keV (pulses at phases $\sim0.3$ and $\sim0.7$); 
with the decreasing of the source luminosity the relative
intensity of a second peak is decreasing and the profile become single-peaked.
The observed pulse profile dependences on the energy and luminosity can be roughly understood in terms of
a simple geometrical model: the angle between the rotation axis
and direction to the observer has a such value that allow us to see one accretion column entirely,
but another column is partially obscured by the neutron star surface.
Thus, the hottest parts of the second column are obscured and the corresponding peak
in the profile should disappear with the growing of the energy.
In terms of this model the decreasing of  the second peak with the luminosity decreasing can be explained
by the decreasing of the accretion column height and consequently by the obscuration of its colder parts.
Naturally for the accurate description of the pulse profile behaviour more complicated models are needed. For example
it is important to know the beam function of the accretion column at different luminosities and energies. Due to the closeness
of radiating areas to the neutron star surface some relativistic effects (e.g., light bending) should be taken into account also.
Thus, the construction of even simple model is very complicated task.
Nevertheless, if the above assumptions are correct  they can give us the possibility to estimate some
basic geometrical parameters of
this X-ray pulsar. 

We did not detect such drastic changes of pulse profile near the cyclotron frequency 
as in case of V0332+53, but the preliminary analysis shown that the profile demonstrates a wavy behaviour with the 
energy: near the main cyclotron
frequency and its harmonics the phase of the first peak in the profile is slightly shifting 
``down''; between CRSF harmonics the phase is shifting ``up'' (Fig. 4). 

\subsection{GX 301-2 and Her X-1}

Two another well known X-ray pulsars with cyclotron features in their spectra, Her X-1 and GX 301-2,
were studied for the searching of the pulse profiles changes with the energy.
The intensity maps of pulse profiles of these pulsars in high intensity states are presented in Fig. 5.
In the left panel the pulse profile of GX 301-2 obtained with
the IBIS telescope onboard the INTEGRAL observatory  in the hard energy band (18-60keV) is shown. 
The Her X-1 
pulse profile obtained with the RXTE observatory
in the 3-40 keV energy band is shown in the right panel.
The preliminary analysis shows that as in the case of 4U0115+63 
no drastic changes of the pulse profile with the energy and near the cyclotron frequency are observed.

\section{Conclusion}

We presented results of the analysis of the \textit{INTEGRAL} and
\textit{RXTE} data obtained for two transient sources,  V0332+53 and 4U0115+63, during the outbursts, 
and two persistent ones, GX 301-2 and Her X-1.
The most important and interesting results are:\\

-- for the first time we studied in detail the evolution of the
cyclotron line energy in the spectrum of V0332+53 with the source luminosity and showed that it is
linearly increasing with the luminosity decreasing in the same
way as the change of the height of the accretion column;

-- the strong pulse profile changes with the luminosity, especially
near the cyclotron line, are revealed for V0332+53;

-- it was shown that for 4U0115+63 the cyclotron line energy depends on the
source luminosity by a complex manner;

-- the possible wavy behaviour of 4U0115+63 pulse profile with the energy is revealed;

-- the preliminary results on GX 301-2 and Her X-1 does not revealed
evident changes of their pulse profiles with the energy.

Detailed studies of the population of pulsars with CRSFs are in a progress.

\section{Acknowledgements}

This work was supported by the Russian Foundation for Basic
Research (project no.04-02-17276), the Russian Academy of Sciences
(The Origins and evolution of stars and galaxies program) and
grant of President of RF (NSh-1100.2006.2). AL acknowledges financial
support from the Russian Science Support Foundation.

\end{document}